# Sodium catalytic effect in the Na$_x$Li$_{6-x}$C$_{60}$ hydrogen storage process

Mattia Gaboardi,[1,2] Chiara Milanese,[3] Giacomo Magnani,[1] Alessandro Girella,[3] Daniele Pontiroli,[1] Mauro Riccò.[1]

1. Dipartimento di Fisica e Scienze della Terra, Università degli Studi di Parma. Parco Area delle Scienze 7/A, I-43124 Parma (Italy).
2. ISIS Facility, Rutherford Appleton Laboratory, Chilton, Didcot, Oxfordshire OX11 0QX, United Kingdom.
3. Pavia Hydrogen Lab, C.S.G.I. & Department of Chemistry - Physical Chemistry Division, University of Pavia. Viale Taramelli 16, I-27100 Pavia (Italy)

We report on the hydrogen sorption investigation of the mixed alkali cluster intercalated fulleride series Na$_x$Li$_{6-x}$C$_{60}$. These compounds are isostructural to Na$_6$C$_{60}$ and Li$_6$C$_{60}$ while the cubic lattice parameter is linearly dependent on $x$. The H$_2$ absorption/desorption was studied by means of charge/discharge kinetic and coupled calorimetric – manometric measurements. By varying the stoichiometry, we found the best compromise between absorption rate, temperature and amount of hydrogen in Na$_1$Li$_5$C$_{60}$. This system is able to reversibly absorb up to 4.3 wt% H$_2$ at 280 °C, which is 70 °C lower in temperature than its parent compound Li$_6$C$_{60}$. Furthermore, the kinetics is improved of 67% with respect Li$_6$C$_{60}$ and the dehydrogenation enthalpy is 13 kJ/mol H$_2$ lower.

## Introduction

Despite the first theoretical studies on lithium and sodium decorated fullerenes,[1–3] depicted as super-fulleroid structures, with several metal ions coordinated by a C$_{60}$; all the structural investigations made up to now demonstrated that, for high stoichiometries, Nature chooses structure in which the metal clusterizes in the large voids of the face centred cubic (*fcc*) structure.[4,5] Consequently, the theoretically predicted hydrogen absorption mechanism (*e.g.*: Kubas interaction)[6] was not in good agreement with the observation of a spillover like effect, found in the experimental studies.[7,8] In this kind of fullerides, hydrogen seems to be dissociated by the presence of partly ionized alkali clusters, made of few atoms (4-9 depending on the stoichiometry). Finally, the high occupancy of the t$_{1g}$-LUMO states of C$_{60}$ (usually charged up to 6 electrons) promotes the hydrogen chemisorption through the formation of a C-H sp$^3$ covalent bond.[9–11] In the recent past, Li$_6$C$_{60}$ and Na$_6$C$_{60}$ were investigated and it was demonstrated that they can reversibly absorb up to 5 and 2.1 wt% H$_2$ respectively[9,12,13] in their bare form and up to 5.9 and ~2.7 wt% H$_2$ when doped with catalysts.[12,14] In particular, Na$_6$C$_{60}$ can absorb 4 wt% H$_2$,[13] with a reversible absorption/desorption between the Na$_6$C$_{60}$H$_{18}$ and Na$_6$C$_{60}$H$_{36}$ species at 375 °C (2.1 wt% H$_2$ stored), and it can be completely dehydrogenated only at 550 °C.[12] Conversely, Li$_6$C$_{60}$ is completely dehydrogenated above 400 °C. Both lithium and sodium intercalated fullerides present advantages and disadvantages: while Li$_6$C$_{60}$ absorbs high amount of hydrogen, its stability in the hydrogenated phase (Li$_6$C$_{60}$H$_y$) is stronger than the hydrogenated Na$_6$C$_{60}$. This causes the onset of desorption to be higher in temperature with respect to the parent Na intercalated phase, although the major desorption event of Li$_6$C$_{60}$H$_y$ occurs at lower temperature than in Na$_6$C$_{60}$H$_y$.[13] In particular, Teprovich *et al.* measured the activation energy ($E_a$) for the 2-steps dehydrogenation process of Na$_6$C$_{60}$H$_y$ and Li$_6$C$_{60}$H$_y$ and found that the sodium intercalated phase present two lower energy barriers ($E_a$~119 and 170 kJ/mol) compared to its lithium counterpart ($E_a$~154 and 190 kJ/mol). Another study carried out by means of coupled manometric – calorimetric measurements concluded that the overall dehydrogenation enthalpy value for Li$_6$C$_{60}$ is about 63 kJ/mol.[14] In the case of Na$_6$C$_{60}$ the enthalpy of reaction for the formation of C$_{60}$H$_{36}$ + 6NaH from Na$_6$C$_{60}$ was predicted to be 56 kJ/mol H$_2$ while in case of Na$_{10}$C$_{60}$ was measured to be 52 kJ/mol H$_2$.[15] The onset temperature for dehydrogenation decreases from 306 °C for Li$_6$C$_{60}$ to about 250 °C for Na$_6$C$_{60}$.[13,14] It is also worth pointing out that the addition of catalysts, useful for improving the kinetics of absorption and the maximum value of absorbed hydrogen, does not affect the enthalpy of desorption.[14] This is in agreement with the role of catalyst, present in form either of micro- and nano-particles, in dissociating the hydrogen molecules, while the desorption of hydrogen from a C-H bond in hydrofullerene mainly depends on the lability of this bond. From this point of view, the transition metal catalyst plays a non-local role (being dispersed in the carbon matrix), while the dehydrogenation of $C_{60}H_y$ is a local process (occurring within the cell). In order to improve not only the absorption kinetics, but also the desorption process; one has to modify the local structure (*i.e.* intercalated ions or clusters, charged state of $C_{60}$, *etc.*). In this paper, we investigate the synthesis and the hydrogen sorption properties of the mixed phases of lithium and sodium intercalated fullerene, $Na_xLi_{6-x}C_{60}$. The aim of this work is to find the best compromise between absorption and kinetics of these materials.

## Methods

Materials were synthesized by following a 2-step procedure. For a typical $Na_xLi_{6-x}C_{60}$ fulleride, about 350 mg of $C_{60}$ (99.9 %, MER Corp.) were ground in an agate mortar with *x* moles of $NaN_3$ (99.99 %, Sigma-Aldrich). Before using, $NaN_3$ was anhydrified by precipitation from ethanol and then treated in dynamic high vacuum at 150 °C for several hours. The powder was pelletized; the pellets placed in tantalum foil bags and then treated in a Pyrex® vial connected to a turbo-molecular vacuum pump. Materials were heated in dynamic high vacuum up to 250 °C with a rate of 60 °C/h and then at 450 °C by ramping at 10 °C/h. At this temperature, the sample was annealed for one day and finally cooled down to room temperature. The as produced $Na_xC_{60}$ was then analysed by means of X-ray diffraction to check the phase. In the second step of synthesis, $Na_xC_{60}$ samples were grinded and mixed with (6-*x*) moles of granular lithium (99 %, Sigma-Aldrich), previously cut in very small flakes. The mixture was milled in an agate ball mill (Fritsch Mini-Mill Pulverisette23, 5 mL volume with 5 agate spheres of 10 mm diameter) at 30 Hz for 60', divided in 6 rounds of 10' followed by 5' of break each. The obtained black powder was pelletized, placed in tantalum bags, within a sealed Pyrex® vial, and treated at 270 °C for 2 days under static high vacuum conditions. Preliminary X-ray powder diffraction was carried out by means of a Bruker D8 Discover instrument (Cu-$K_{\alpha 1}$ radiation), working in Debye-Scherrer geometry and equipped with an area detector (GADDS). Sealed glass capillaries were filled with powder and the measurement was performed, while spinning, collecting data for several hours per frame. Hydrogen absorption investigations were performed on the as prepared samples in a PCTPro-2000 manometric instrument (Setaram). About 300 mg of sample was heated from room temperature up to 280 °C at 5 °C/min under 100 bar of hydrogen and a 10 h isotherm was appended at the end of the ramp. Hydrogen desorption kinetic measurements were performed by heating the sample at 400 °C under 0.5 bar of hydrogen and appending 10 h of isotherm. Coupled calorimetric – manometric measurements were performed by connecting the high-pressure stainless steel cell of the Sensys high pressure DSC by Setaram with the PCTPro. About 30 mg of the hydrogenated samples (after the first charging run) were discharged by heating from room temperature up to 450 °C at 0.5 bar of hydrogen in dynamic mode (heating rate = 5 °C/min). All the operations of synthesis and handling of materials were carried out in air and moisture free conditions by operating in vacuum or within an Ar glove box ($O_2$ and $H_2O$ levels <1 ppm).

## Results and discussion

The synthesis of $Na_xC_{60}$ phases by means of the azide method produces well-crystallized samples, as shown in powder X-ray diffractions of Fig.SI1. The Le Bail analysis of those spectra confirmed the formation of the well-known *fcc* phases of $Na_2C_{60}$ (for x = 1, 2, and 3) and $Na_6C_{60}$ (for x = 6),[16] and the monoclinic phase of $Na_4C_{60}$ (x = 4 and 5).[17] The $Na_xLi_{6-x}C_{60}$ phases exhibit broadened peaks (see Figure 1). These are in part due to the reduced size of powders and the increased disorder caused by the high-energy ball milling. On the other hand, a decrease of the *fcc* symmetry, inducing a pseudo-*fcc* arrangement of fullerenes, is commonly observed in Li containing fullerides, leading to *fcc* peaks splitting. In case of low resolution (such as for X-ray powder diffraction) the convolution of these peaks can be confused as peak broadening. The refinement of the pseudocubic lattice parameter by Le Bail analysis of the diffractograms, adopting the $Fm\overline{3}m$ cell of $C_{60}$, demonstrated an increasing trend, according to the relative increase of *x*. For x = 6, we found *a* = 14.37 Å, in good agreement with the literature.[16] Satellite peaks near the 111 reflection, at ~10.5°, and the peak at 18-20°, between the 220 and the 311 reflections, are normally attributed to hexagonal distortion of the *fcc* lattice, as stacking fault type of defects [18]. Due to the *fcc* (or pseudo-*fcc*) arrangements of $C_{60}$ molecules, the only way to fill the free space with 6 atoms of Na (Li) is allowing the formation of an alkaline cluster in the central octahedral void of the cell. This cluster is tetrahedral in case of $Na_6C_{60}$ [16] and not yet characterized in the novel $Na_xLi_{6-x}C_{60}$ phases.

As well as previously reported, the addition of Na is known to destabilize the hydrogen sorption process, promoting the dissociation of hydrofullerene below 300 °C.[10,15] Thus, we decided to hydrogenate the samples below this temperature. The hydrogenation was carried out at 280 °C under 100 bar $H_2$. The first hydrogen absorption and desorption cycle is reported, as a function of time, in Figure 2. Hydrogen absorption – desorption cycles showed that both the gravimetric capacity and the absorption rate significantly improve by decreasing the Na content and the values obtained for the mixed compounds $Na_{0.5}Li_5C_{60}$, $Na_1Li_5C_{60}$, $Na_2Li_4C_{60}$, and $Na_3Li_3C_{60}$ are better than those for pure $Li_6C_{60}$, testifying the catalytic activity of Na when added in small amount to the Li-fullerite. The hydrogen sorption reaction is reversible for all the stoichiometries and the cycling ability is good in the frame of 3 cycles (maximum tested up to now).

The dehydrogenation calorimetric profiles recorded on the samples after the first hydrogenation run show an increasing in complexity by increasing the Na content. For $Na_1Li_5C_{60}$ one only and quite sharp endothermic peak centred at about 310 °C is present, resembling the signal of pure $Li_6C_{60}$. By increasing the Na content, one endothermic shoulder at higher temperature appears, making the peak broader and broader and shifting its maximum temperature up to 370 °C for the $Na_5LiC_{60}$ sample. This large peak resembles the highest temperature signal of $Na_6C_{60}$. Other two endothermic peaks compare in the calorimetric trace of this compound (to our best knowledge the first obtained in literature), one centred at 170 °C and the other at 260 °C. Both these signals are coupled with mass loss, hence they are dehydrogenation steps. The desorption enthalpy under the highest temperature signal decrease from 66 kJ/mol $H_2$ for pure $Na_6C_{60}$ to 50 kJ/mol $H_2$ for $Na_1Li_5C_{60}$, value lower than that determined by us for $Li_6C_{60}$ (63). The most absorbing system is the one with *x*=0.5 stoichiometry, showing 4.7 wt% $H_2$ reversibly desorbed. Anyway, from the combined analysis of these data and of the calorimetric curves is evident that $Na_1Li_5C_{60}$ is the most promising stoichiometry also considering thermodynamics, an important technological parameter. Its maximum absorption reaches 4.3 wt% $H_2$, when $Li_6C_{60}$ only get 2.7 wt% $H_2$ after a longer isotherm landing at 280 °C. The most important results on the hydrogen storage parameters of $Na_xLi_{6-x}C_{60}$ are reported in Table 1.

*Table 1: kinetic absorption properties (Absorbed $H_2$, onset temperature and rate) and $\Delta H_{des}$ of the $Na_xLi_{6-x}C_{60}$ phases.*

| | Absorption capacity at 280 °C (wt% $H_2$) | Absorption capacities after isotherm (wt% $H_2$) | $T_{onset}$ desorption (°C) | Absorption Rate ($10^{-2}$wt%/min) | $\Delta H_{des}$ (kJ/mol) |
|---|---|---|---|---|---|
| $Li_6C_{60}$ | 1.8 | 2.8 | 306 | 3.9 | 63 |
| $Na_{0.5}Li_{5.5}C_{60}$ | 3.4 | 4.9 | 280 | 7.0 | / |
| $Na_1Li_5C_{60}$ | 3.4 | 4.3 | 260 | 6.5 | 50 |
| $Na_2Li_4C_{60}$ | 3.3 | 3.9 | 250 | 5.9 | / |
| $Na_3Li_3C_{60}$ | 2.7 | 3.2 | 200 | 4.7 | / |
| $Na_4Li_2C_{60}$ | 1.9 | 2.4 | 200 | 3.3 | / |
| $Na_5LiC_{60}$ | 1.2 | 2.3 | 200 | 3.5 | 64 |
| $Na_6C_{60}$ | 1.5 | 1.8 | 160 | 2.3 | 66 |

Figure 3 displays the DSC of hydrogenated $Na_1Li_5C_{60}$ under discharge at 400 °C, showing three exothermic desorptions and an overall dehydrogenation enthalpy of 50 kJ/mol $H_2$, which is 13 kJ/mol lower than the dehydrogenation enthalpy of hydrogenated $Li_6C_{60}$.

The X-ray powder diffraction patterns of hydrogenated $Na_xLi_{6-x}C_{60}$ are reported in Figure 4 for $x$ = 0, 0.5 and 2. After hydrogenation at 280 °C, the *fcc* cell results expanded because of the $C_{60}$ volume increasing due to C-H bonds formation. In case of $Li_6C_{60}$ the lattice evolved from *fcc* to *bcc* when hydrogenated.[9] Anyway, a small substitution of Li with Na seems to be enough to overcome this structural transition, which has been recognized as a kinetics limiting process.[19] It is worth to highlight the formation of LiH and NaH, although a quantitative estimation cannot be done from XRD data, due to the unknown structure of the hydrofullerene anion and the low scattering factor of Li and H.

The analysis of the time derivative of desorption curves allows to better separate the different processes involved when varying the stoichiometry. In Figure 5 the rates of desorption are reported as a function of time and the profile of desorption has been fit to Gaussian functions. It is possible to identify at least four main processes. An isolated process occurring at around 150 °C (appreciable only for $x$>0.5) and three convoluted processes occurring above 200 °C. The partial dehydrogenations extracted from the fits are shown in Figure 6.**Errore. L'origine riferimento non è stata trovata.** It is evident that the first peak of desorption, between 140 - 155 °C depending on $x$, is highly affected by the Na content, being absent for $Li_6C_{60}$ and $Na_{0.5}Li_{5.5}C_{60}$, slightly visible in $Na_1Li_5C_{60}$, and progressively more consistent for 2≤$x$≤6. It is worth to pointing out that the percentage of desorbed hydrogen is not only dependent on the amount of hydrogen but also on the weight of the absorber. Therefore, an increasing (decreasing) wt% of $H_2$, when varying the amount of Na in the structure, does not necessarily corresponds to the same variation in the stoichiometric content of hydrogen. For instance, assuming that the only product of hydrogenation is $Na_xLi_{6-x}C_{60}H_y$, a 1 wt% $H_2$ in $Na_1Li_5C_{60}$ and $Na_6C_{60}$ corresponds to a different value of y = 7.8 and 8.6 respectively. The amount of hydrogen released in the first process is very low and varies between 0.02 (for x=1) and 0.2 wt% $H_2$ (in $Na_6C_{60}$), corresponding to about 0-2 hydrogens per $C_{60}$, while the minimum temperature for this step is found for $x$=3 (140 °C). The second process is visibly lower in temperature for Na containing samples (247 °C for $x$=2) than for $Li_6C_{60}$ (291 °C) and the amount of desorbed hydrogen varies from 0.3 ($Li_6C_{60}$) to 0.95 wt% ($x$=4), corresponding to 2-8 hydrogens per $C_{60}$, depending on $x$. The third process, occurring between 308 and 347 °C, is the most important and dependent on $x$. In the case of $Li_6C_{60}$ it occurs at 323 °C and only 0.3 wt% $H_2$ is released (about 2 hydrogens per $C_{60}$). In the mixed phases, the maximum desorption is reached for x = 0.5 at 330 °C, corresponding to 3.3 wt% $H_2$ (about 26 hydrogens per $C_{60}$) and decreases progressively with $x$. The last process is the highest in temperature; it decreases in temperature with $x$ varying from 380 °C ($Li_6C_{60}$) to 350 °C ($x$=0.5) and for some sample is completed during the isotherm. In the case of $Li_6C_{60}$ this is the most important process, coinciding to 2 wt% $H_2$ desorbed (about 15 hydrogens). Anyway, for $x$≠0, the amount of desorbed hydrogens ranges from 8 to 12 (depending on $x$), corresponding to about 0.3-1.2 wt% $H_2$. The same analysis has been attempted on the absorption data, anyway, the peaks were not well separated as for desorption and their deconvolutions led to ambiguous results.

Considering the amount of hydrogen desorbed per single step, it is possible to reconstruct the complete dehydrogenation processes of $Na_xLi_{6-x}C_{60}$ compounds. It was found that the hydrogenation involves the segregation of part of the metal in form of hydride (NaH and LiH in our case).[10,11] Therefore, we can adopt the following general equation:

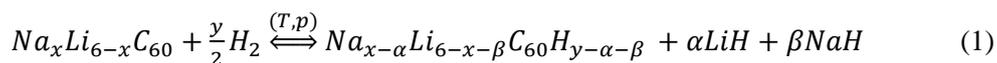
$$Na_xLi_{6-x}C_{60} + \frac{y}{2}H_2 \stackrel{(T,p)}{\Longleftrightarrow} Na_{x-\alpha}Li_{6-x-\beta}C_{60}H_{y-\alpha-\beta} + \alpha LiH + \beta NaH \qquad (1)$$

Here $\alpha$ and $\beta$ are the amount of lithium and sodium hydrides respectively. Unfortunately, is difficult to quantify these values from XRD data, since the structure of hydrofulleride is unknown and a quantitative phase analysis via Rietveld refinement is not possible. In $Na_{10}C_{60}$, $Li_6C_{60}$, and $Li_{12}C_{60}$ part of the alkali metal is de-intercalated in form of hydride during the absorption process.[10,11,19] The measurements on hydrogenated $Na_xLi_{6-x}C_{60}$ highlighted the presence of both LiH and NaH (see Figure 4). It is possible give an estimation of $y$ assuming $\alpha$=$\beta$=0. Another assumption we have done is that hydrofullerene composition is $C_{60}H_{2n}$, where n≥1, since only an even number of hydrogens is considered to produce a stable $C_{60}H_y$ molecule.[20] Below, for simplicity, we will exclude from discussion the first absorption process at 140 °C. For $Li_6C_{60}$ about 20 hydrogens are chemisorbed on $C_{60}$ at the end of the absorption process. During the first two dehydrogenation steps at 291 and 324 °C about 0.28 wt% $H_2$ per step are released (2 hydrogens per step: $C_{60}H_{18}$ and $C_{60}H_{16}$ are respectively formed). The third dehydrogenation process allows to desorb 2 wt% $H_2$, corresponding to the remaining hydrogens. For x=0.5, the total chemisorbed hydrogen (4.74 wt%) corresponds to $C_{60}H_{38}$. During the following three steps of dehydrogenation at 280, 331 and 350 °C, $C_{60}H_{34}$, $C_{60}H_{24}$ and $C_{60}$ are respectively formed. When x=1, about 36 hydrogens are chemisorbed on fullerene ($C_{60}H_{36}$) at the end of the absorption process. During the second dehydrogenation process (257 °C) $C_{60}H_{36}$ loses about 4 hydrogens per molecule. The third step at 315 °C brings to $C_{60}H_{22}$ and the complete dehydrogenation is achieved at 370 °C. The path of dehydrogenation for x=2 is $C_{60}H_{30} \rightarrow C_{60}H_{28}$ (247 °C) $\rightarrow C_{60}H_{20}$ (308 °C) $\rightarrow C_{60}$ (370 °C). For x=3: $C_{60}H_{26} \rightarrow C_{60}H_{20}$ (261 °C) $\rightarrow C_{60}H_2$ (310 °C) $\rightarrow C_{60}$ (360 °C). For x=4: $C_{60}H_{22} \rightarrow C_{60}H_{12}$ (280 °C) $\rightarrow C_{60}H_{10}$ (322 °C) $\rightarrow C_{60}$ (377 °C). For x=5: $C_{60}H_{20} \rightarrow C_{60}H_{12}$ (269 °C) $\rightarrow C_{60}H_{10}$ (317 °C) $\rightarrow C_{60}$ (356 °C). Finally, for $Na_6C_{60}$: $C_{60}H_{14} \rightarrow C_{60}H_8$ (270 °C) $\rightarrow C_{60}$ (353 °C).

The first stage around 140 °C was considered together with the first of the three high temperature processes in the calculation of the dehydrogenation paths, due to the low value of hydrogen involved (0-2 hydrogens per $C_{60}H_y$, depending on x). Since this process is very far in temperature (about 100 °C) from the other three processes, we attribute it to extrinsic hydrogen species (*e.g.*: hydrogen not bound to carbon). Comparing the four stages of desorption, it is clear that the first step only occurs when Na is intercalated. A possible explanation is that those hydrogens responsible for this process are likely to form a chemical bond with sodium, either in form of ion (*e.g.*: NaH), or clustered with intercalated lithium. The second hypothesis is likely to occur when, as previously assumed[21], the mechanism of hydrogenation can be explained through a spillover-like effect. In fact, the hydrogen molecule is quickly dissociated on the metal cluster during the first stage of absorption, until the cluster become less effective to perform this task. Then, the hydrogenation of fullerene continues at lower rate since alkali cluster has been partly deintercalated in form of hydride. This was already evidenced in Pt-Pd doped $Li_6C_{60}$, where the presence of a catalyser allowed to continue the fast process at the limit, even when the LiH was segregated.[14] Anyway, in that case the presence of the catalyser did not significantly influenced the desorption enthalpy. On the contrary, by substituting Li with small fractions of Na, the absorption process is faster. Apparently, Na, more than Li, is likely to remain intercalated in the hydrofulleride structure (as also suggested by the absence of NaH XRD peaks in hydrogenated $Na_{0.5}Li_{5.5}C_{60}$ and $Na_1Li_5C_{60}$, see Figure 4). The presence of Na in the hydrofulleride cell may affect the $C_{60}H_y$ state of charge, thus the C-H bond strength and the dehydrogenation enthalpy.

## Conclusions

Mixed alkali-cluster intercalated Na and Li fullerides $Na_xLi_{6-x}C_{60}$ have been synthesized by means of two step procedure consisting in the thermal decomposition of sodium azide in $C_{60}$ and the ball-milling of metallic lithium with $Na_xC_{60}$. The hydrogen storage investigation proved the catalytic effect of Na in promoting the hydrogenation of $C_{60}$ anion. On one hand, the weight of hydrogen chemisorbed by $Na_{0.5}Li_{5.5}C_{60}$, and $Na_1Li_5C_{60}$, is improved by 74, and 50%, respectively, with respect to $Li_6C_{60}$, with about 67% better rate of absorption. On the other hand, the DSC coupled manometric measurements evidenced that also the dehydrogenation enthalpy is affected when Li is substituted by small amounts of of Na. In particular, $\Delta H_{des}$ = 50 kJ/mol $H_2$ for hydrogenated $Na_1Li_5C_{60}$, 13 kJ/mol less than hydrogenated $Li_6C_{60}$. This study allowed us to establish that $Na_1Li_5C_{60}$ represents the best compromise between the amount of stored hydrogen, kinetics of ab/des-sorption, and enthalpy of dehydrogenation.

## Acknowledgements
This work has received support from the Cariplo foundation (Project number 2013-0592, "Carbon based nanostructures for innovative hydrogen storage systems") and from the European Union's Horizon 2020 research and innovation programme under the Marie Skłodowska-Curie grant agreement No 665593.

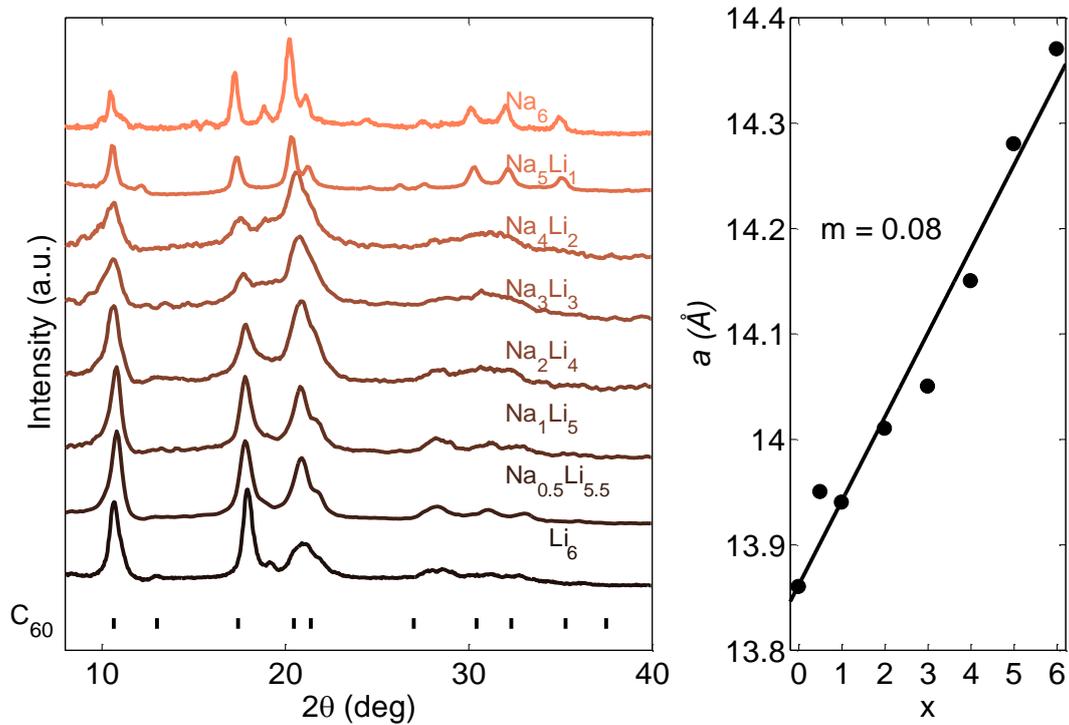

*Figure 1. Left: X-ray powder diffractions of $Na_xLi_{6-x}C_{60}$. Right: fcc lattice parameter as a function of Na content from 0 to 6, as obtained from Le Bail analysis of the diffraction patterns.*

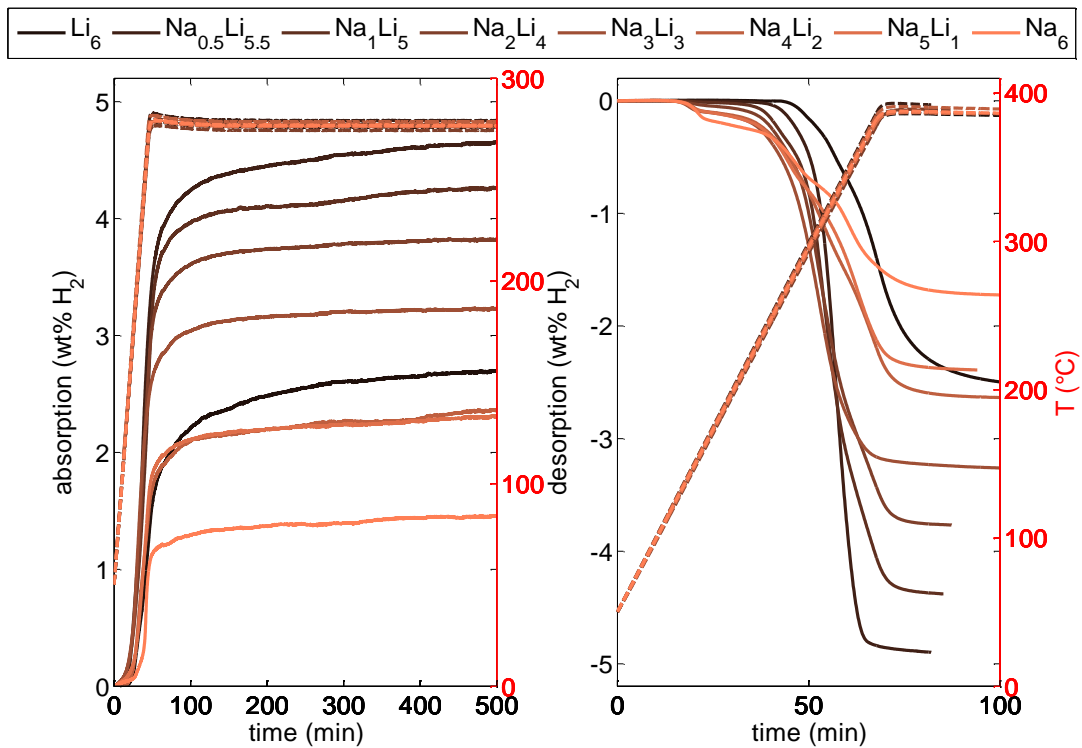

*Figure 2. 1st cycle of hydrogen absorption (left) and desorption (right) of Na$_x$Li$_{6-x}$C$_{60}$ are reported as a function of time (solid lines, left Y-axis). The right Y-axes represent the temperature (dashed lines).*

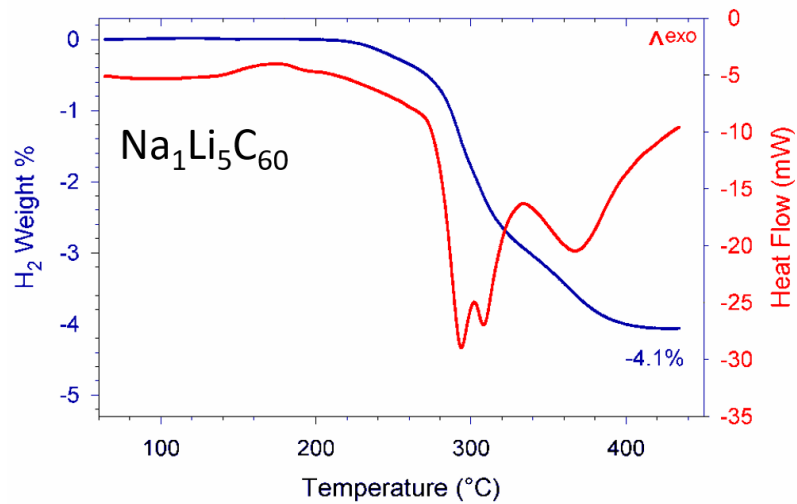

*Figure 3. Differential scanning calorimetry of Na$_1$Li$_5$C$_{60}$, showing three exothermic desorption peaks with a global enthalpy of 50 kJ/mol H$_2$.*

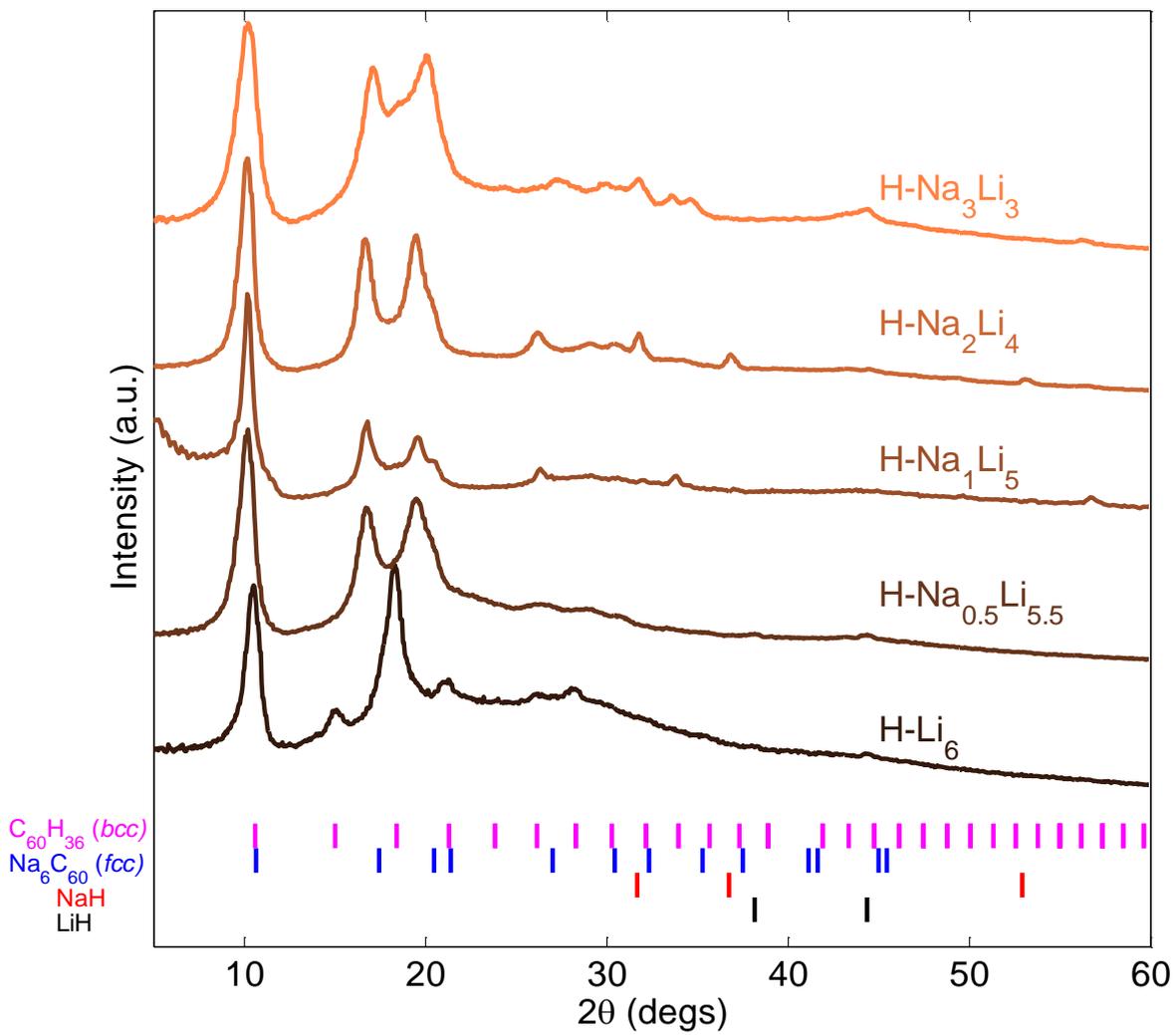

*Figure 4. X-ray powder diffraction patterns of selected $Na_xLi_{6-x}C_{60}$ after the first hydrogen absorption at 280 °C (100 bar). The pattern of $Li_6C_{60}$ hydrogenated at 350 °C (100 bar) is also reported for comparison.*

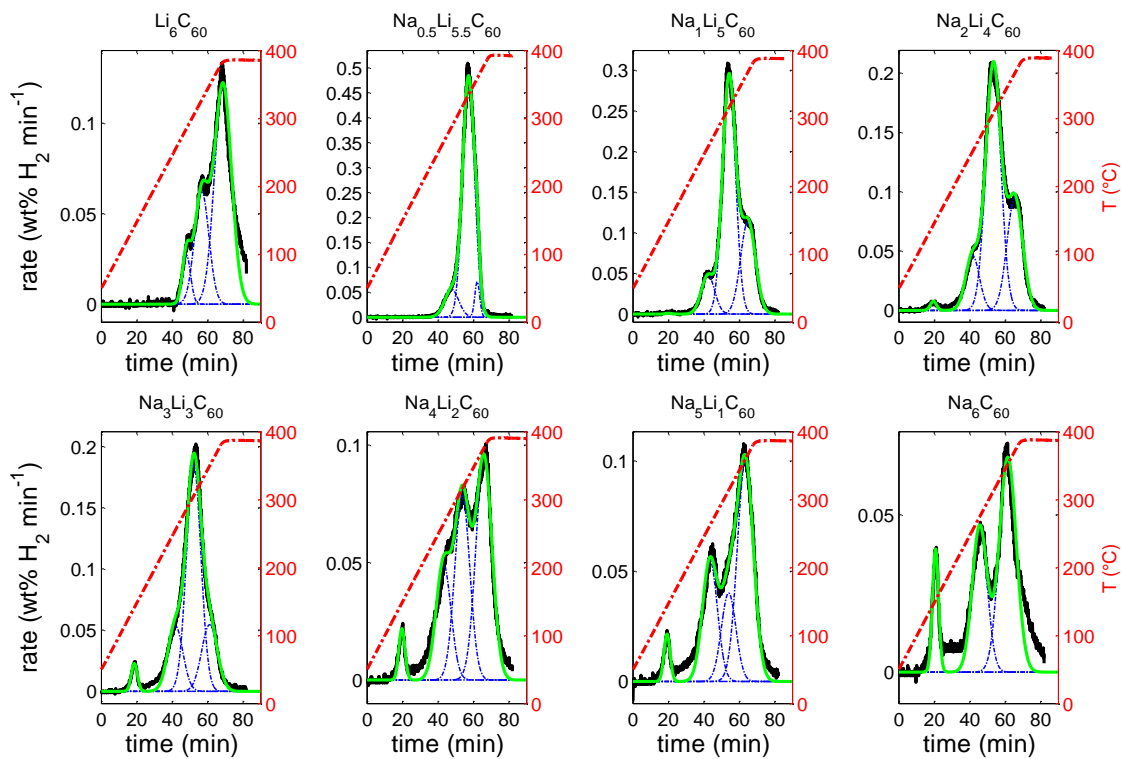

*Figure 5. hydrogen desorption rates (black) and fit (green) according to Gaussian functions (blue) of hydrogenated $Na_xLi_{6-x}C_{60}$ reported as a function of time (left Y-axes, red dashed lines).*

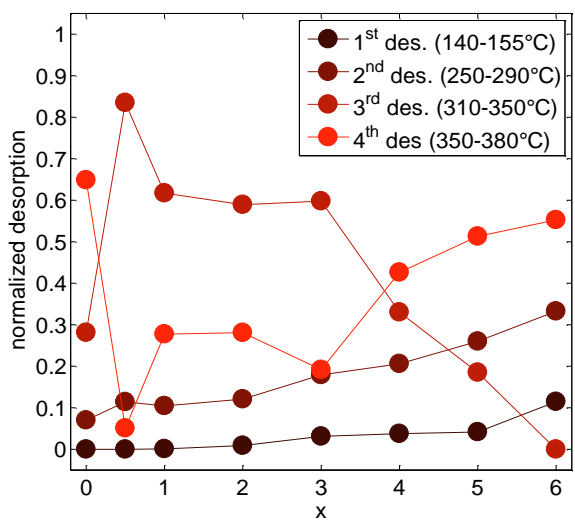

*Figure 6. $Na_xLi_{6-x}C_{60}H_y$ normalized partial desorptions reported as a function of x.*

# References

1. Kohanoff, J., Andreoni, W. & Parrinello, M. A possible new highly stable fulleride cluster: Li12C60. *Chem. Phys. Lett.* **198,** 472–477 (1992).

2. Chandrakumar, K. R. S. & Ghosh, S. K. Alkali-metal-induced enhancement of hydrogen adsorption in C60 fullerene: an ab Initio study. *Nano Lett.* **8,** 13–9 (2008).

3. Sun, Q., Jena, P., Wang, Q. & Marquez, M. First-principles study of hydrogen storage on Li12C60. *J. Am. Chem. Soc.* **128,** 9741–5 (2006).

4. Yildirim, T. *et al.* Intercalation of sodium heteroclusters into the C60 lattice. *Nature* **360,** 568–571 (1992).

5. Giglio, F. *et al.* Li12C60: A lithium clusters intercalated fulleride. *Chem. Phys. Lett.* **609,** 155–160 (2014).

6. Kubas, G. J., Ryan, R. R., Swanson, B. I., Vergamini, P. J. & Wasserman, H. J. Characterization of the first examples of isolable molecular hydrogen complexes, M(CO)3(PR3)2(H2) (M = molybdenum or tungsten; R = Cy or isopropyl). Evidence for a side-on bonded dihydrogen ligand. *J. Am. Chem. Soc.* **106,** 451–452 (1984).

7. Aramini, M. *et al.* Muon spin relaxation reveals the hydrogen storage mechanism in light alkali metal fullerides. *Carbon N. Y.* **67,** 92–97 (2014).

8. Gaboardi, M. *et al.* Hydrogen storage mechanism and lithium dynamics in Li12C60 investigated by μSR. *Carbon N. Y.* **90,** 130–137 (2015).

9. Teprovich, J. A. *et al.* Synthesis and characterization of a lithium-doped fullerane (Lix-C60-Hy) for reversible hydrogen storage. *Nano Lett.* **12,** 582–9 (2012).

10. Mauron, P. *et al.* Reversible hydrogen absorption in sodium intercalated fullerenes. *Int. J. Hydrogen Energy* **37,** 14307–14314 (2012).

11. Mauron, P. *et al.* Hydrogen Sorption in Li12C60. *J. Phys. Chem. C* **117,** 22598–22602 (2013).

12. Knight, D. a *et al.* Synthesis, characterization, and reversible hydrogen sorption study of sodium-doped fullerene. *Nanotechnology* **24,** 455601 (2013).

13. Teprovich, J. a., Knight, D. a., Peters, B. & Zidan, R. Comparative study of reversible hydrogen storage in alkali-doped fulleranes. *J. Alloys Compd.* **580,** S364–S367 (2013).

14. Aramini, M. *et al.* Addition of transition metals to lithium intercalated fullerides enhances hydrogen storage properties. *Int. J. Hydrogen Energy* **39,** 2124–2131 (2014).

15. Mauron, P. *et al.* Hydrogen Desorption Kinetics in Metal Intercalated Fullerides. *J. Phys. Chem. C* **119,** 1714–1719 (2015).

16. Rosseinsky, M. J. *et al.* Structural and electronic properties of sodium-intercalated C60. *Nature* **356,** 416–418 (1992).

17. Kubozono, Y. *et al.* Structure and physical properties of Na4C60 under ambient and high pressures. *Phys. Rev. B* **63,** 45418 (2001).

18. Céolin, R. *et al.* A new hexagonal phase of fullerene C60. *Chem. Phys. Lett.* **314,** 21–26 (1999).

19. Gaboardi, M. *et al.* In Situ Neutron Powder Diffraction of Li6C60 for Hydrogen Storage. *J. Phys. Chem. C* **119,** 19715–19721 (2015).

20. Cahill, P. A. in *The Chemistry of Fullerenes* 53–66 (1995). doi:10.1142/9789812386250_0004

21. Aramini, M. *et al.* Muon spin relaxation reveals the hydrogen storage mechanism in light alkali metal fullerides (SI). *Carbon N. Y.* 1–2 (2013). doi:10.1016/j.carbon.2013.09.063